\documentclass[12pt,journal, onecolumn, draftcls]{IEEEtran}
\usepackage{amsmath}
\usepackage{epsfig}
\usepackage{amssymb}
\usepackage{graphicx}
\usepackage{hyperref}
\usepackage{color}
\usepackage{amsmath,amssymb,epsfig,bbm,amsbsy,cite}
\usepackage{amsmath}
\usepackage{amssymb}

\begin{document}
\title{Power Allocation Based on SEP Minimization in Two-Hop
Decode-and-Forward Relay Networks}
\author{Arash~Khabbazibasmenj,~\IEEEmembership{Student
Member,~IEEE} \ and \ Sergiy A.~Vorobyov, \IEEEmembership{Senior
Member,~IEEE}
\thanks{This work is supported in parts by the Natural Science and
Engineering Research Council (NSERC) of Canada and the Alberta
Ingenuity Foundation, Alberta, Canada. The authors are with the
ECE Dept., University of Alberta, 9107-116 St., Edmonton, Alberta,
T6G~2V4 Canada. Emails: {\tt \{khabbazi,
vorobyov\}@ece.ualberta.ca.}

{\bf Corresponding author:} Sergiy A.~Vorobyov, Dept. Elect. and
Comp. Eng., University of Alberta, 9107-116 St., Edmonton,
Alberta, T6G 2V4, Canada; Phone: +1 780 492 9702, Fax: +1 780 492
1811. Email: {\tt vorobyov@ece.ualberta.ca}. }}
\maketitle
%
%
\begin{abstract}
The problem of optimal power allocation among the relays in a
two-hop decode-and-forward cooperative relay network with
independent Rayleigh fading channels is considered. It is assumed
that only the relays that decode the source message correctly
contribute in data transmission. Moreover, only the knowledge of
statistical channel state information is available. A new simple
closed-form expression for the average symbol error probability is
derived. Based on this expression, a new power allocation method
that minimizes the average symbol error probability and takes into
account the constraints on the total average power of all the
relay nodes and maximum instant power of each relay node is
developed. The corresponding optimization problem is shown to be a
convex problem that can be solved using interior point methods.
However, an approximate closed-form solution is obtained and shown
to be practically more appealing due to significant complexity
reduction. The accuracy of the approximation is discussed.
Moreover, the so obtained closed-form solution gives additional
insights into the optimal power allocation problem. Simulation
results confirm the improved performance of the proposed power
allocation scheme as compared to other schemes.
\end{abstract}


\section{Introduction}
\label{sec:intro} Cooperative relay networks enjoy the advantages
of the multiple-input multiple-output (MIMO) systems such as, for
example, high data rate and low probability of outage by
exploiting the inherent spatial diversity without applying
multiple antennas at the nodes. In cooperative relay networks,
after receiving the source message, relay nodes process and then
retransmit it to the destination. Different cooperation protocols
such as decode-and-forward (DF), amplify-and-forward (AF), coded
cooperation, and compress-and-forward can be used for processing
the message at the relay nodes \cite{IEEEhowto:Survey},
\cite{IEEEhowto:Lane}. The benefits of cooperative relay networks
can be further exploited by optimal power allocation among the
source and relay nodes. Specifically, based on the knowledge of
the channel state information (CSI) at the relays and/or
destination, the system performance can be improved by optimally
allocating the available power resources among the relays
\cite{IEEEhowto:Survey}--\cite{Sergiy}.

Different power allocation schemes have been proposed in the
literature \cite{Yener}--\cite{Arash}. These schemes differ from
each other due to the different considerations on the network
topology, assumptions on the available CSI, use of different
cooperation protocols for relay nodes, and use of different
performance criteria \cite{IEEEhowto:Survey}. Most of the existing
power allocation methods require the knowledge of instantaneous
CSI to enable optimal power distribution \cite{IEEEhowto:Survey},
\cite{Yener}--\cite{Maric}. The application of such methods is
practically limited due to the significant amount of feedback
needed for transmitting the estimated channel coefficients and/or
the power levels of different nodes. This overhead problem becomes
even more severe when the rate of change of the channel fading
coefficients is fast.

In this paper, we aim at avoiding the overhead problem by
considering only statistical CSI which is easy to obtain.
Recently, some power allocation methods based on statistical CSI
have been proposed\cite{IEEEhowto:Luo}--\cite{Sou}. The optimal
power allocation problem among multiple AF relay nodes that
minimizes the total power given a required symbol error
probability (SEP) at the destination is studied in \cite{Maham}.
The problem of optimal power distribution in a three node DF relay
network which aims at minimizing the average SEP is studied in
\cite{Sadek1}. The authors of \cite{Sadek2} study the power
allocation problem in a multi-relay DF cooperative network in
which the relay nodes cooperate and each relay coherently combines
the signals received from previous relays in addition to the
signal received from the source to minimize the average SEP. The
power allocation problem aiming at minimizing the average SEP in a
cooperative network consisting of two DF relay nodes in
Nakamgi-$m$ fading channel has been studied in \cite{Sou}. All of
the aforementioned power allocation methods are based on
minimizing or bounding the asymptotic approximate average SEP
which is valid at high signal-to-noise ratios (SNRs) and is not
applicable at low and moderate SNRs.

In our initial conference contribution \cite{Arash}, a power
allocation method for multi-relay DF cooperative network with
Rayleigh fading channels that minimizes the exact average SEP has
been proposed. However, the assumption of correct decoding in
relay nodes used in \cite{Arash} limits its practical
applicability. For obtaining a more practically suitable power
allocation method, we consider in this paper the case when relay
nodes may not be able to decode the source signal correctly. We
derive the optimal power allocation in a multi node DF relay
network with Rayliegh fading channels in which only relays which
have decoded the source message correctly contribute in the data
relaying. More specifically, after receiving the source message,
only the relays which decode the message correctly retransmit it
to the destination. A new exact and simple closed-form expression
for average SEP is derived. Then a new power allocation strategy
is developed by minimizing the exact average SEP rather than its
high SNR approximation under the constraints on the total average
power of all relays and maximum powers of individual relays. Only
the knowledge of the average channel gains, i.e., the knowledge of
the variances of the channel coefficients, is assumed to be
available. We show that the corresponding optimization problem is
convex and, thus, can be solved using the well established
interior point methods. In order to find better insights into the
power allocation problem, we derive an approximate closed-form
solution to the problem and discuss the accuracy of the
approximation used. We also show by simulations that the exact
numerical and approximate solutions provide close average SEP
performance and that the proposed power allocation scheme
outperforms other schemes.

The paper is organized as follows. System model is introduced in
Section~II. A simple closed-form expression for the average SEP is
derived in Section~III, while power allocation (exact and
approximate) strategies based on SEP minimization are derived in
Section IV. Simulation results are given in Section V followed by
conclusion. All technical proofs are given in Appendix. This paper
is reproducible research \cite{RepResearch} and the software
needed to generate the simulation results will be made available
together with the paper. It can be also requested by the reviewers
if needed.

\section{System Model}
\begin{figure}[t]
\begin{center}
\includegraphics[scale=.5]{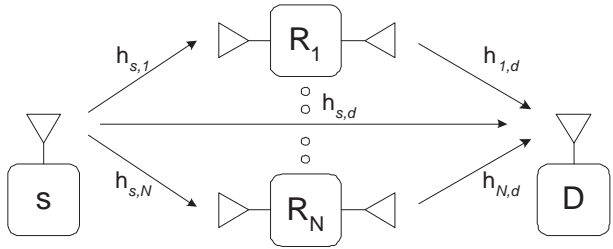}
\caption{System Model: A single source communicates with a single
destination through $N$ relay nodes and a direct path.}
\label{Model}
\end{center}
\end{figure}

\label{sec:format} Consider a wireless relay network with a single
source communicating with a single destination through $N$ relay
nodes as it is shown in Fig.~\ref{Model}. The relays are assumed
to be half-duplex, that is, the relays either transmit or receive
the signal at the same frequency at any given time instant.
Therefore, every data transmission from the source to the
destination occurs in two phases. In the first phase, the source
node transmits its message to the destination and the relay nodes,
while in the second phase, relay nodes retransmit the source
message to the destination. The channels between the source and
the relay nodes, between the relay nodes and the destination, and
the direct path are assumed to be flat Rayligh fading and
independent from each other. The source and relay nodes use the
$M$-phase shift keying (M-PSK) modulation\footnote{Note that other
types of modulation can be straightforwardly adopted and PSK
modulation is considered only because of space limitation.} for
data transmission where $M$ is the size of the constellation.
Relay nodes use DF cooperation protocol for processing the
received signal from the source node. The received signal at the
destination and the $i$th relay node in the first phase can be
expressed, respectively, as
\begin{eqnarray}
y_{s,d}&=&\sqrt{p_0} h_{s,d} x + n_{s,d} \\
y_{s,i}&=&\sqrt{p_0} h_{s,i} x + n_{s,i}, \ \ \ \ \ i=1,\ldots,N
\label{rece}
\end{eqnarray}
where $x$ is the source message of unit power, $p_0$ is the
transmit power of the source node, $h_{s,d}$ and $h_{s,i}$ denote
the channel coefficients between the source and the destination
and between the source and the $i$th relay node, respectively, and
$n_{s,d}$ and $n_{s,i}$ are the complex additive white Gaussian
noises (AWGNs) in the destination and in the $i$th relay,
respectively. Since the channel fading is Rayleigh distributed,
the channel coefficients are modeled as independent complex
Gaussian random variables with zero mean and variances $m_{s,d}$
and $m_{s,i}$, respectively, for the channels between the source
and the destination and between the source and the $i$th relay
node. The additive noises are zero mean and have variance $N_0$.

After decoding the received signal from the source \eqref{rece},
only the relay nodes which have decoded the source message
correctly retransmit it to the destination through orthogonal
channels using time division multiple access (TDMA) or frequency
division multiple access (FDMA). By means of an ideal cyclic
redundancy check (CRC) code applied to the transmitted information
from the source, relays can determine whether they have decoded
the received signal correctly or not \cite{Sadek1}. Then the
probability that $i$th relay node can decode the received signal
correctly conditioned on the instantaneous CSI can be expressed as
\cite{IEEEhowto:Alouini}
\begin{equation}
\label{CD} \alpha_i(h_{s,i}) =1- \frac{1}{\pi}
\int_0^{\frac{M-1}{M}\pi} e^{{-\frac{g_{PSK} } {\sin^2(\theta)
N_0}}|h_{s,i}|^2 p_0} \, d\theta
\end{equation}
where $g_{PSK} \triangleq \sin^2 (\pi/M)$.

Let $\boldsymbol \phi = ( \phi_1, \phi_2, \ldots, \phi_N)^T$
denote a vector that indicates whether each relay has decoded the
source message correctly or not. Specifically, $\boldsymbol
\phi(i) = 1$, if $i$th relay node decodes the source message
correctly, and $\boldsymbol \phi(i) = 0$, otherwise. Here $( \cdot
)^T$ stands for the transpose. In the rest of the paper, we refer
to the vector $\boldsymbol \phi$ as the vector of decoding state
at the relay nodes. Since $\boldsymbol \phi$ consists of only
binary values, there are in total $2^N$ different combinations
that the vector $\boldsymbol \phi$ can take. Moreover, there is a
one-to-one correspondence between the binary representation of
decimal numbers $k = 0, \ldots, 2^N-1$ and different values that
the vector $\boldsymbol \phi$ can take. For example, in a
cooperative network with two relays, if the relay enumerated as
first decodes the source message correctly and the relay
enumerated as second decodes it incorrectly, the corresponding
decoding state vector is $(1, 0)^T$ and the corresponding
representation in decimal is $1 \cdot 2^1 + 0 \cdot 2^0 = 2$. For
simplicity, we represent hereafter each possible combination of
vector $\boldsymbol \phi$ by its corresponding decimal number $k$
and denote this combination as $\boldsymbol \phi_k$. For example,
$\boldsymbol \phi_0$ corresponds to the situation when all relay
nodes decode the source message incorrectly.

The received signal from the $i$th relay node which is able to
decode the source message correctly, that is, $\boldsymbol \phi
(i) = 1$, at the destination can be modeled as
\begin{equation}
y_{i,d} = \sqrt{p_i} h_{i,d}x + n_{i,d}
\end{equation}
where $p_i$ is the transmitted power of the $i$th relay node,
$h_{i,d}$ is the channel coefficient between the $i$th relay node
and the destination, and $n_{i,d}$ is the AWGN with zero mean and
variance $N_0$. The channel coefficient $h_{i,d}$ is modeled as an
independent complex Gaussian random variable with zero mean and
variance $m_{i,d}$ due to the Rayleigh fading assumption. It is
worth stressing that the assumption that the channel coefficients
are independent from each other is applicable for relay networks
because the distances between different relay nodes are typically
large enough. It is assumed that the destination knows perfectly
the instantaneous CSI from the relays to the destination and the
instantaneous CSI of the direct link. The knowledge of
instantaneous CSI for the links between the source and the relay
nodes is not needed. Then the maximal ratio combining (MRC)
principle can be used at the destination to combine received
signals form the source and relay nodes. As a result of MRC, the
received SNR at the destination conditioned on the decoding state
at the relay nodes, i.e., $\boldsymbol \phi_k$, can be expressed
as
\begin{equation}
\label{RSNR} \gamma_D (\boldsymbol \phi_k) = \gamma_{s} +
\sum_{\{i | \boldsymbol \phi_k (i) = 1\}} \gamma_{i}
\end{equation}
where $\gamma_s \triangleq p_0 |h_{s,d}|^2/N_0$ and $\gamma_i
\triangleq p_i |h_{i,d}|^2/N_0$ are the received SNRs from the
source and the $i$th relay node at the destination, respectively.
Here $\gamma_s$ and $\gamma_i$, $i=1, \cdots, N$ are exponential
random variables with means $p_0 m_{s,d}/N_0$ and $p_i
m_{i,d}/N_0$, respectively. Moreover, $\gamma_s$ and $\gamma_i$,
$i=1, \cdots, N$ are all statically independent.

\section{Average SEP } \label{sec:typestyle}
For the considered case when the data transmission is performed
using the M-PSK modulation, the SEP of the signal at the
destination conditioned on the channel states $CS
=\{h_{s,d},h_{s,i},h_{i,d}\}$ and the decoding state at the relay
nodes $\boldsymbol \phi_k$  can be written as
\cite{IEEEhowto:Alouini}
\begin{eqnarray} \label{Pe1}  P_{e} \{ CS , \boldsymbol \phi_k \}
\!\!\!\!\!&=&\!\!\!\!\! \frac{1}{\pi} \int_0^{\frac{M-1}{M}\pi}
e^{-\frac{g_{PSK} } {\sin^2(\theta)} \gamma_D(\boldsymbol \phi_k)}
\,d \theta .
\end{eqnarray}
Using the total probability rule, the SEP conditioned on the
channel states can be expressed as
\begin{equation} \label{Pe4}
P_{e} \{ CS \} = \sum_{k=0}^{2^N-1} P_r \{
\boldsymbol \phi_k\} P_{e} \{ CS, \boldsymbol \phi_k \}
\end{equation}
where $P_r \{ \boldsymbol \phi_k \}$ is the probability of the
decoding state $\boldsymbol \phi_k$ that can be calculated as
\begin{equation}
P_r \{ {\boldsymbol \phi}_k \} = \prod_{\{ i | {\boldsymbol
\phi}_k (i) = 1 \}} \alpha_i \cdot \prod_{\{ i | {\boldsymbol
\phi}_k (i)=0\}} (1 - \alpha_i) \label{ProbS}
\end{equation}
where $\alpha_i$ is the probability of correct decoding in $i$th
relay node \eqref{CD}. Note that for obtaining \eqref{ProbS}, the
independency of the AWGNs at the relay nodes has been exploited.

The average SEP 
can be obtained by averaging (\ref{Pe4}) over $h_{s,d}$,
$h_{s,i}$, $i=1, \cdots, N$, and $h_{i,d}$, $i=1, \cdots, N$ and
using the fact that $P_r \{ \boldsymbol \phi_k \}$ is
statistically independent from $P_{e} \{ CS, \boldsymbol \phi_k
\}$. The latter follows from the statistical independence between
the channel coefficients and the fact that $P_r \{ \boldsymbol
\phi_k \}$ depends only on $\{ h_{s,i} \}_{i=1}^N$ and $P_{e} \{
CS, \boldsymbol \phi_k \}$ depends only on $\{ h_{i,d} \}_{i=1}^N$
and $h_{s,d}$. Then the average SEP can be expressed as
\begin{equation} \label{Peee}
P_e \!=\! \frac{1}{\pi} \sum_{k=0}^{2^N-1} E \left\{ P_r \{
\boldsymbol \phi_k\} \right\} E \left\{ \int_0^{\frac{M-1}{M}\pi}
\! e^{-\frac{g_{PSK} } {\sin^2(\theta)}\gamma_D
(\boldsymbol\phi_k)} \, d\theta \right\}
\end{equation}
where $E \{ \cdot \}$ denotes the expectation operation.

The second expectation in \eqref{Peee} can be computed as
\begin{eqnarray} \label{rightside}
\!\!\!\!\!& &\!\!\!\!\! E \left\{ \int_0^{\frac{M-1}{M}\pi }
e^{-\frac{g_{PSK} }{\sin^2(\theta)} \gamma_D ( \boldsymbol
\phi_k) } \, d\theta \right\} \nonumber  \\
\!\!\!\!\!&=&\!\!\!\!\! \int_0^{\frac{M-1}{M}\pi} \! E \left\{
e^{-\frac{g_{PSK} } {\sin^2( \theta) N_0} \left( \sum\limits_{\{i
| \boldsymbol \phi_k(i)=1\}} p_i |h_{i,d}|^2 + p_0 |h_{s,d}|^2
\right) } \right\} \, d\theta \nonumber \\
\!\!\!\!\!&=&\!\!\!\!\! \int_0^{\frac{M-1}{M}\pi} \prod_{ \{i |
\boldsymbol \phi_k (i) = 1\} \cup \{i=0\} } \! \frac{\sin^2
(\theta)}{\sin^2 (\theta) \!+\! b_i p_i } \, d\theta
\end{eqnarray}
where $b_i \triangleq g_{PSK} m_{i,d}/N_0$, $i=1, \ldots, N$ and
$b_0 \triangleq g_{PSK} m_{s,d}/N_0$. Similarly, the first
expectation in \eqref{Peee} can be computed as
\begin{eqnarray} \label{leftside}
E \left\{ Pr \{ \boldsymbol \phi_k\} \right\} \!\!&=&\!\! E
\left\{ \prod_{\{i|\boldsymbol \phi_k(i)=1\}} \! \alpha_i \cdot
\prod_{\{i|\boldsymbol \Phi_k (i)=0\}} ( 1 - \alpha_i) \right\}
\nonumber \\
\!\!&=&\!\! \prod_{\{i|\boldsymbol \Phi_k(i)=1\}} \beta_i \cdot
\prod_{\{i|\boldsymbol \Phi_k(i)=0\}} (1 - \beta_i)
\end{eqnarray}
where
\begin{eqnarray} \label{betaiinit}
\beta_i \!\!&=&\!\!  E \left\{ 1- \frac{1}{\pi}
\int_0^{\frac{M-1}{M}\pi} e^{{-\frac{g_{PSK} } {\sin^2(\theta)
N_0}}|h_{s,i}|^2 p_0} \, d\theta \right\} \nonumber \\
\!\!&=&\!\! 1 - \frac{1}{\pi} \int_0^{\frac{M-1}{M}\pi}
\frac{\sin^2(\theta)}{\sin^2(\theta) + c_i p_0} \, d\theta
\end{eqnarray}
and $c_i \triangleq g_{PSK} m_{s,i}/{N_0}$, $i=1,\ldots,N$.

Substituting \eqref{rightside} and \eqref{leftside} in
\eqref{Peee}, the average SEP can be equivalently expressed as
\begin{eqnarray} \label{Pe5}
P_{e} \!\!&=&\!\! \frac{1}{\pi} \sum_{k=0}^{2^N-1} \left(
\prod_{\{i|\boldsymbol \phi_k (i)=1\}} \beta_i \cdot
\prod_{\{i|\boldsymbol \phi_k (i)=0\}} (1-\beta_i) \right.
\nonumber \\
\!\!&\times&\!\! \left. \int_0^{\frac{M-1}{M}\pi} \!\!\!
\prod_{\{i|\boldsymbol \phi_k (i)=1\}\cup \{i=0\}}
\frac{\sin^2(\theta)}{\sin^2(\theta) + b_i p_i } \, d\theta
\right) . \label{sep_1}
\end{eqnarray}
Setting $\beta_0=1$ for notation simplicity, the average SEP
expression in \eqref{sep_1} can be simplified as
\begin{eqnarray}
P_{e} \!\!&=&\!\! \frac{1}{\pi} \sum_{k=0}^{2^N-1} \left(
\prod_{\{i | \boldsymbol \phi_k (i)=0\}} (1 - \beta_i) \prod_{\{i
| \boldsymbol \phi_k (i)=1\} \cup \{i=0\}} \beta_i
\right. \nonumber \\
\!\!&\times&\!\! \left. \int_0^{\frac{M-1}{M}\pi} \!\!\!
\prod_{\{i|\boldsymbol \Phi_k (i)=1\}\cup \{i=0\}}
\frac{\sin^2(\theta)}{\sin^2(\theta) + b_i p_i } \, d\theta
\right) \label{NEWsep_1} \\
\!\!&=&\!\! \frac{1}{\pi} \int_0^{\frac{M-1}{M} \pi} \left(
\sum_{k=0}^{2^N-1} \prod_{\{i | \boldsymbol \phi_k (i)=0\}}
(1 - \beta_i) \right. \nonumber \\
\!\!&\times&\!\! \left. \prod_{\{i|\boldsymbol \phi_k (i)=1\}
\cup\{i=0\}} \beta_i \cdot \frac{ \sin^2 (\theta)}{\sin^2(\theta)
+ b_i p_i } \, d\theta \right). \label{sep_2}
\end{eqnarray}

Finally, the expression \eqref{sep_2} can be further simplified as
\begin{equation} \label{sep_4}
P_{e} = \frac{1}{\pi} \int_0^{\frac{M-1}{M}\pi} \prod_{i=0}^{N}
\left( (1 - \beta_i)\!+\! \frac{\beta_i
\sin^2(\theta)}{\sin^2(\theta) + b_i p_i } \right)  \, d\theta.
\end{equation}
To verify the latter result, note that \eqref{sep_2} can be
obtained by simply expanding \eqref{sep_4}. Moreover, after
finding the integral in \eqref{betaiinit}, $\beta_i$,
$i=1,\ldots,N$ can be expressed as
\begin{eqnarray}
\beta_i \!\!&=&\!\! 1- \! \sqrt{\frac{c_i p_0 }{ c_i p_0 \!+\! 1}}
\!{ \left(\! \frac{\tan^{-1} \left( \! \cot \left( \!
\frac{M-1}{M}\pi \right)\sqrt{\frac{c_i p_0 }{c_i p_0 \!+\! 1}}
\right)}{\pi} \!-\!
\frac{1}{2} \! \right)} \nonumber \\
\!\!&-&\!\! \left( \frac{1}{2} \!-\! \frac{\tan^{-1} \left (\cot
\left( \frac{M-1}{M} \pi \right) \right)} {\pi} \right), \;
i=1,\ldots,N. \label{closed-form2}
\end{eqnarray}
Here $\beta_i$ has a meaning of statistical average of the correct
decoding probability in the $i$th relay node \eqref{CD} with
respect to the channel coefficients. It is worth noting that the
average SEP expression \eqref{sep_4} is used later for finding the
optimal power distribution among the relay nodes.

Using partial fraction decomposition, it is possible to further
simplify \eqref{sep_4} and find a closed-form expression for the
average SEP. It is worth noting that for simplicity and because of
space limitations, we hereafter assume that $b_i \neq b_j$, $i
\neq j$, $i, j=0, \ldots, N$. However, if the condition $b_i \neq
b_j$, $i \neq j$, $i, j = 0,\ldots,N$ does not hold, the following
average SEP derivation approach remains unchanged, while the only
change is the need of using another form of the partial fraction
decomposition. Thus, it is straightforward to derive closed-form
expression for the average SEP in the case when $b_i \neq b_j$, $i
\neq j$, $i, j = 0,\ldots,N$ does not hold by using the same steps
as we show next.

Toward this end, we first rewrite the integral
inside \eqref{NEWsep_1}, denoted hereafter as $I$, by changing the
variable as
\begin{eqnarray}
\!\!\!\!\!I\!\!&=&\!\!\!\!\! \int_0^{\frac{M-1}{M}\pi} \prod_{\{i
|\boldsymbol \phi_k(i)=1\} \cup \{i=0\}}
\frac{\sin^2(\theta)}{\sin^2(\theta)
+ b_i p_i } \, d\theta \nonumber \\
\!\!\!\!\!&=&\!\!\!\!\! \int_{{\rm cot} \left( \frac{M-1}{M}\pi
\right)}^{\infty} \! \frac{1}{1 \!+\! x^2} \!\! \prod_{\{i
|\boldsymbol \phi_k(i)=1\}\cup \{i=0\}} \!\!\!
\frac{1}{ (1 \!+\! b_i p_i) \!+\! b_i p_i x^2 } \, dx . \nonumber \\
\!\!& &\!\! \label{EQ2}
\end{eqnarray}
Then applying the partial fraction technique to the right-hand
side of \eqref{EQ2} and using the fact that $b_i \neq b_j$, $i\neq
j$, $i, j=0,\ldots,N$, we obtain
\begin{eqnarray}
I \!\!&=&\!\! \int_{{\rm cot} \left( \frac{M-1}{M} \pi
\right)}^{\infty} \frac{1}{1+x^2} \! \nonumber\\
\!\!&-&\!\!\!\!\!\! \sum_{\{i |\boldsymbol \phi_k(i)=1\} \cup
\{i=0\}} \!\!\!\!\! \frac{ \prod_{(\{j|\boldsymbol
\phi_k(j)=1\}\cup\{j=0\})-\{ j\neq i\}} \frac{1}{1 \!-\! \frac{b_j
p_j}{b_i p_i} }}{ (1 \!+\! \frac{1}{b_i p_i})+x^2 } \, dx .
\nonumber \\
\!\!& &\!\! \label{inint}
\end{eqnarray}
The integral in \eqref{inint} is the summation of the terms that all
have the form of $1/(a+x^2), \ a > 0$. By using the fact that
$\int 1/(a+x^2) dx =1/\sqrt{a} \tan^{-1}(x/ \sqrt{a}), \ a > 0$,
\eqref{inint} can be calculated as
\begin{eqnarray}
I \!\!&=&\!\!\!\!\!\!\!\!\! \sum_{\{\!i |\boldsymbol \phi_k(i) =
1\} \cup \{\!i=0\}} \!\!\!\Bigg( \!\!{ \Bigg(
{\tan^{-1}\!\!\left(\! \cot \left( \! \frac{M\!-\!1}{M}\pi
\!\!\right) \!\!\sqrt{\frac{b_i p_i }{b_i p_i \!+\! 1}} \right)}
\!-\! \frac{\pi}{2} \! \Bigg)}
\nonumber \\
\!\!&\times&\!\! \sqrt{\frac{b_i p_i}{b_i p_i \!+\! 1}}
\prod_{(\{j|\boldsymbol \phi_k(j)=1\}\cup\{j=0\})-\{ j\neq i\}}
\frac{1} {1-\frac{b_j p_j }{b_i p_i}} \Bigg) \nonumber \\
\!\!&+&\!\! \left( \frac{\pi}{2}- {\tan^{-1} (\cot
(\frac{M-1}{M}\pi ) )} \right)   \label{inintnew}
\end{eqnarray}

Substituting the so obtained expression \eqref{inintnew} in
\eqref{NEWsep_1} and also expanding the resulted expression, the
average SEP can be equivalently expressed as
\begin{eqnarray}  \label{eq:closedform}
P_e \!\!\!&=&\!\!\! \sum_{k=0}^{2^N-1} \!  \prod_{\{i|
\phi_k(i)=1\} \cup \{i=0\}} \!\!\!\! \beta_i \cdot \prod_{\{i|
\phi_k(i)=0\}}^N (1-\beta_i)  \nonumber \\
\!\!\!&\times&\!\!\! \sum_{\{\!i| \phi_k(i)=1\!\} \cup \{\!i=0\}}
\! \Bigg( \!\!{ \Bigg( \frac{\tan^{-1} (  \cot ( \!
\frac{M-1}{M}\pi ) \sqrt{\frac{b_i p_i }{b_i p_i \!+\! 1}} )}{\pi}
\!-\! \frac{1}{2} \Bigg)} \nonumber \\
\!\!\!&\times&\!\!\! \sqrt{\frac{b_i p_i}{b_i p_i \!+\! 1}}
\prod_{(\{\!j|\boldsymbol \phi_k (j) = 1\!\}\cup\{\!j=0\})-\{\!j\neq
i\!\}} \frac{1} {1 \!-\! \frac{b_j p_j}{b_i p_i}} \Bigg) \nonumber \\
\!\!\!&+&\!\!\! \sum_{k=0}^{2^N-1} \!\! \prod_{\{\!i| \phi_k
(i)=1\!\} \cup \{\!i=0\!\}} \beta_i \times \prod_{\{i| \phi_k
(i)=0\} }^N (1-\beta_i) \nonumber  \\
\!\!\!&\times&\!\!\! \left( \frac{1}{2}- \frac{\tan^{-1} \left
(\cot \left(\frac{M-1}{M}\pi \right) \right)}{\pi} \right).
\label{EQclosed1}
\end{eqnarray}
Moreover, we can modify the first term of \eqref{EQclosed1} by
taking the product $ \prod_{\{i|\phi_k(i)=1\} \cup \{i=0\}}
\beta_i$ into the summation part of that term and expanding the
product such that $\beta_i$ is multiplied by the corresponding
term in the summation and also modify the second term of
\eqref{EQclosed1} by using the following equality which is easy to
prove
\begin{equation}
\sum_{k=0}^{2^N-1} \!\! \prod_{\{\!i| \phi_k (i)=1\!\} \cup
\{\!i=0\!\}} \beta_i \cdot \prod_{\{i| \phi_k (i)=0\} }^N
(1-\beta_i)=1 .
\end{equation}
By doing so, the average SEP can be rewritten as
\begin{eqnarray}
P_e \!\!\!&=&\!\!\! \sum_{k=0}^{2^N-1} \! \prod_{\{i|\Phi_k(i)
\!=\! 0\}} (1\!-\!\beta_i) \nonumber \\
\!\!\!&\times&\!\!\! \sum_{\{\!i| \phi_k(i)\!=\!1\!\} \cup
\{\!i=0\}} \! \Bigg( \!{ \beta_i \Bigg( \frac{\tan^{-1} ( \cot (
\! \frac{M-1}{M} \pi ) \sqrt{\frac{b_i p_i}{b_i p_i \!+\! 1}} )}
{\pi} \!-\! \frac{1}{2} \Bigg)} \nonumber  \\
\!\!\!&\times&\!\!\! \sqrt{\frac{b_i p_i}{b_i p_i \!+\! 1}}
\prod_{(\{\!j|\boldsymbol \Phi_k (j) \!=\!
1\!\}\cup\{\!j=0\!\})-\{\!j\neq i\!\}} \frac{\beta_j} {1 \!-\!
\frac{b_j p_j }{b_i p_i}} \Bigg) \nonumber \\
\!\!\!&+&\!\!\! \left( \frac{1}{2} \!-\! \frac{\tan^{-1} \left(
\cot \left(\frac{M-1}{M}\pi \right) \right)}{\pi} \right).
\label{eq1}
\end{eqnarray}

Finally, after rearranging and factoring out the terms, we can
obtain the following closed form expression for the average SEP
\begin{eqnarray}
P_e \!\!\!&=&\!\!\!\! \sum_{i=0}^N \!\Bigg( \!{ \beta_i \Bigg(
\frac{\tan^{-1} (\cot ( \! \frac{M-1}{M}\pi ) \sqrt{\frac{b_i p_i
}{b_i p_i \!+\! 1}} )}{\pi} \!-\! \frac{1}{2} \Bigg)}
\sqrt{\frac{b_i p_i}{b_i p_i \!+\! 1}} \nonumber \\
\!\!\!&\times&\!\!\! \prod_{\{\!j|j=0,\ldots,N\}-\{\! j\neq i\!\}}
\left( \frac{\beta_j}{1 \!-\!\frac{b_j p_j }{b_i p_i}} \!+\! (1
\!-\! \beta_j) \right) \Bigg) \nonumber  \\
\!\!\!&+&\!\!\! \left( \frac{1}{2} \!-\! \frac{\tan^{-1} \left
(\cot \left(\frac{M-1}{M}\pi \right) \right)}{\pi} \right).
\label{eq2}
\end{eqnarray}
Note that \eqref{eq1} can be obtained by simply expanding
\eqref{eq2}. It is interesting to mention that if we consider
similar conditions as in \cite{Arash}, in which all relays are
capable of decoding the source message correctly, i.e., $\beta_i
=1, i=1, \ldots, N$, and also there is no direct link between
source and destination, i.e., $b_0 = 0$, the closed-form average
SEP \eqref{eq2} simplifies to the one that was obtained in
\cite{Arash}. The closed-form expression \eqref{eq2} is simple and
does not include any other functions rather than basic analytic
functions.

\section{Power allocation}
In this section, we address in details the problem of optimal
power allocation among the relay nodes such that the average SEP
is minimized. Only statistical information on the channel states
is used. Moreover, we assume that the power of the source node
$p_0$ is fixed and, in turn, the average probabilities of the
correct decoding of the relays, i.e., $\beta_i$, $i = 1, \ldots,
N$, are also fixed. The relevant figure of merit for the
performance of relay network is then the average SEP that is
derived above in closed form. It enables us to apply power
allocation also in the case when the rate of change of channel
fading is high. Since only statistical CSI is available and relay
nodes retransmit only if they decode the source message correctly,
the averaged power of the relay nodes over CSI and probability of
correct decoding need to be considered. It is easy to see that the
average power used by the $i$th relay node equals $\beta_i p_i$.
Indeed, since $\beta_i$ is the statistical average of the
probability of correct decoding in the $i$th relay node, the
transmitted power of the $i$th relay node $p_i$ weighted by
$\beta_i$ gives average power used by the $i$th relay node during
a single transmission.

In the following, we develop a power allocation strategy by
minimizing the average SEP \eqref{sep_4}, while satisfying the
constraint on the total average power used in relay nodes $p_R$
and the constraints on maximum instantaneous powers per every
relay $p_i^{max}$, $i = 1, \ldots, N$. With the knowledge of the
average channel gains $m_{s,i} = {E} \{ |h_{s,i}|^2 \}$,
$m_{i,d}={E} \{ |h_{i,d}|^2 \}$, $m_{s,d}={E} \{ |h_{s,d}|^2 \}$,
and specifications on $p_R$, $p_i^{max}$, $i=1, \ldots, N$, and
$p_0$, the power allocation problem can be formulated as
\begin{eqnarray}
& & \min_{\mathbf{p}} \ P_e (\mathbf{p}) \label{obj} \\
& & {\rm s.t.} \quad \sum_{i=1}^N \beta_i {p_i} = p_R \label{equality} \\
& & \qquad \; 0\le p_i\le p_i^{max}, i=1, \ldots, N .
\label{eq:problem}
\end{eqnarray}
where $P_e(\mathbf{p})$ is the average SEP \eqref{sep_4} written
as a function of powers at the relay nodes $\mathbf p = (p_1,
\ldots , p_N)^T$. For notation simplicity, we use hereafter the
following equation for the average SEP in \eqref{sep_4}
\begin{equation}
P_{e}(\mathbf p) = \frac{1}{\pi} \int_0^{\frac{M-1}{M}\pi} g
(\theta, \mathbf{p}) \,d\theta \label{SEP3}
\end{equation}
where
\begin{equation}
g (\theta, \mathbf{p}) \triangleq \prod_{i=0}^{N} \bigg((1 - \beta_i) +
\beta_i \frac{\sin^2(\theta)}{\sin^2(\theta) + b_i p_i }\bigg).
\end{equation}

Note that the optimization problem (\ref{obj})--(\ref{eq:problem})
is infeasible if $\sum_{i=1}^N \beta_i p_i^{max} < p_R$ and it has
the trivial solution, that is, $p_i = p_i^{max}$, if $\sum_{i=1}^N
\beta_i p_i^{max} = p_R$. Thus, it is assumed that $\sum_{i=1}^N
\beta_i p_i^{max} > p_R$. The following theorem about convexity of
the optimization problem (\ref{obj})--(\ref{eq:problem}) is in
order.

\textbf{Theorem~1:} \textit{The optimization problem
(\ref{obj})--(\ref{eq:problem}) is  convex.}

See the proof in Appendix A.

Although this problem does not have a simple closed-form solution,
an accurate approximate closed-form solution can be found as it is
shown in the rest of the paper.  A numerical algorithm for finding
the exact solution of the optimization problem
(\ref{obj})--(\ref{eq:problem}) based on the interior-point
methods is summarized in Appendix B.  Despite the higher
complexity of the numerical method as compared to the approximate
closed-form solution, numerical method can provide an exact
solution, which can be used as a benchmark to evaluate the
accuracy of the approximate solution.


\subsection{Approximate Closed-Form Solution}
\label{ssec:subhead} The optimization problem
(\ref{obj})--(\ref{eq:problem}) is strictly feasible because
$\sum_{i=1}^N \beta_i p_i^{max} > p_R$ as it has been assumed
earlier. Thus, the Slater's condition holds and since the problem
is convex, the Karush-Kuhn-Tucker (KKT) conditions are the
necessary and sufficient optimality conditions
\cite{IEEEhowto:Boyd}. Indeed, since $\sum_{i=1}^N \beta_i
p_i^{max} > p_R$, then $p_i = p_i^{max} p_R / (\sum_{i=1}^N
\beta_i p_i^{max})$ is a strictly feasible point for the
optimization problem \eqref{obj}--\eqref{eq:problem}.

Let us introduce the Lagrangian
\begin{eqnarray}
L(\mathbf{p},\boldsymbol{\lambda},\boldsymbol{\gamma},\nu)
\!\!\!&=&\!\!\!\!\! \frac{1}{\pi} \! \int_0^{\frac{M \! -\!
1}{M}\pi} g (\theta, \mathbf{p}) \, d\theta \!-\!
\sum_{i=1}^N\lambda_i p_i
\nonumber \\
+ \!\!\!\!\!\!&& \!\!\!\!\!\!\!\!\!  \sum_{i=1}^N \gamma_i(p_i
\!-\! p_i^{max}) \!+\! \nu \!\! \left(  \sum_{i=1}^N {\beta_i p_i}
\!-\! p_R \!\! \right) \label{Lagrangian}
\end{eqnarray}
where ${\boldsymbol\lambda} \triangleq (\lambda_1, \cdots,
\lambda_N)^T$ and $\boldsymbol{\gamma} \triangleq (\gamma_1,
\cdots, \gamma_N)^T$ are $N \times 1$ vectors of non-negative
Lagrange multipliers associated with the inequality constraints
$p_i \geq 0$, $i=1, \cdots ,N$ and $p_i \le p_i^{max}$, $i=1,
\cdots ,N$, respectively, and $\nu$ is the Lagrange multiplier
associated with the equality constraint $\sum_{i=1}^N {p_i} =
p_R$. Then the KKT conditions can be obtained as
\begin{eqnarray}
& &\lambda_i \geq 0, \ \ \ \ \  \gamma_i \geq 0, \;  \ \ \ \
\; i=1, \cdots, N \label{KKT1} \\
& & \lambda_ip_i=0, \quad \gamma_i(p_i-p_i^{max})=0, \; i=1,
\cdots, N \\
& & \frac{1}{\pi}\!\! \int_0^{\frac{M-1}{M}\pi}
\!\!\!\!\!\!\!\!\!\!\! \frac{-\beta_i b_i \sin^2(\theta)  g
(\theta, \mathbf{p})}{\sin^4 (\theta) \!+\! (2-\beta_i) b_i
p_i\sin^2(\theta) \!+\! (1-\beta_i)b_i^2p_i^2 } \,
d\theta \nonumber \\
& & \qquad + \nu \beta_i - \lambda_i +\gamma_i \!=\! 0, \qquad
i=1,
\cdots, N \label{eq:lagrange1} \\
& & 0\le p_i\le p_i^{max}, \ \ \ \ \ i=1, \cdots, N   \\
& & \sum_{i=1}^N {\beta_i p_i} = p_R. \label{eq:equality}
\end{eqnarray}

Although the exact optimal solution of the problem
\eqref{obj}--\eqref{eq:problem} can be found through solving the
system \eqref{KKT1}--\eqref{eq:equality} numerically or as it is
summarized in Appendix~B by solving the original problem directly,
a near optimum closed-form solution for the system
\eqref{KKT1}--\eqref{eq:equality}, and thus, optimization problem
\eqref{obj}--\eqref{eq:problem} can be found by approximating the
gradient of the Lagrangian, that is, the left hand side of
\eqref{eq:lagrange1}. Specifically, it can be verified that for
fixed $p_i$, $i=1,\ldots, N$ the function $g(\theta, \mathbf p)$
is strictly increasing/decreasing with respect to $\theta$ in the
intervals $(0,{\pi}/{2})$ and $({\pi} /{2},{(M-1)\pi}/{M})$,
correspondingly. Under the condition that the number of relays is
large enough, the slope of the increment and decrement in the
aforementioned intervals is high and the Chebyshev-type
approximation on the conditions \eqref{eq:lagrange1} is highly
accurate. This approximation of the conditions
\eqref{eq:lagrange1} is of the form
\begin{eqnarray}
\label{aprox} -\frac{\beta_i b_i}{\pi(1 \!+\! (2-\beta_i) b_i p_i
\!+\! (1-\beta_i)b_i^2p_i^2)}  \nonumber  \\
\times  \int^{\frac{M-1}{M}\pi }_0 \!\!\!\!\! g (\theta, \mathbf{p})
d\theta  + \beta_i \nu - {\lambda }_i + {\gamma }_i = 0, & &
\nonumber \\ & & \!\!\!\!\!\!\!\!\!\!\!\!\!\!
\!\!\!\!\!\!\!\!\!\!\!\!\!\! i=1, \cdots, N \label{aprcond}
\end{eqnarray}
where the fact that $\sin^2(\pi/2) = 1$ is used in the ratio
\begin{equation}
\frac{-\beta_i b_i \sin^2(\theta)}{\sin^4 (\theta) \!+\! (2 -
\beta_i) b_i p_i\sin^2(\theta) \!+\! (1-\beta_i)b_i^2p_i^2 }
\label{Approx}
\end{equation}
By rearranging the denominator of \eqref{Approx}, substituting the
approximation \eqref{aprcond} in \eqref{eq:lagrange1}, dividing
the equations (\ref{KKT1})--(\ref{eq:lagrange1}) by the positive
quantity $\int_0^{\frac{M-1}{M}\pi} g(\theta,\mathbf p )
\,d\theta$, and also multiplying these equations by $\pi$, we
obtain
\begin{eqnarray}
& &\lambda_i^\prime \geq 0, \ \ \ \ \  \gamma_i^\prime \geq 0, \;
\ \ \ \  \; i=1, \cdots, N \label{NewNNKKT1} \\
& & \lambda_i^\prime p_i=0, \quad
\gamma_i^\prime(p_i-p_i^{max})=0, \; i=1,
\cdots, N \\
& & -\frac{\beta_i b_i}{\beta_i
(1\!+\!b_ip_i)\!+\!(1\!-\!\beta_i)(1\!+\!b_i p_i)^2}\! +\! \beta_i
{\nu
}^\prime\!-\!\lambda_i^{\prime}\!+\!{{\gamma }_i}^\prime\!\!=\!0
\nonumber \\
& &  \qquad\qquad\qquad\qquad \qquad\qquad \qquad i=1, \cdots, N
\label{NewNNeq:lagrange1} \\
& & 0\le p_i\le p_i^{max}, \ \ \ \ \ i=1, \cdots, N   \\
& & \sum_{i=1}^N {\beta_i p_i} = p_R. \label{NewNNeq:equality}
\end{eqnarray}
where ${\nu }^\prime \triangleq \nu \pi / \int_0^{\frac{M-1}
{M}\pi} g(\theta ) \,d\theta$, ${\gamma }_i^\prime \triangleq
{\gamma }_i \pi / \int_0^{\frac{M-1}{M}\pi} g(\theta ) \,d\theta$,
and \ ${\lambda }_i^\prime$ $\triangleq {\lambda }_i \pi /
\int_0^{\frac{M-1}{M}\pi} g(\theta ) \,d\theta$, $i=1, \cdots, N$.
It is possible to eliminate $\lambda_i^{\prime}$ from the set of
equations \eqref{NewNNKKT1}--\eqref{NewNNeq:equality} in order to
find a simpler set of smaller number of equations. By doing so,
the approximate KKT conditions can be equivalently rewritten as

\begin{eqnarray}
& & \label{eq:NKKT1} {{\gamma }_i}^\prime\ge 0, \quad i=1, \cdots, N \\
& & \label{eq:NKKT2}{{\gamma }_i}^\prime\left(p_i- {p_i}^{max}
\right)=0,
\qquad i=1, \cdots, N \\
& & \label{eq:NKKT5} p_i\left( \beta_i {\nu
}^\prime\!+\!{{\gamma }_i}^\prime -\frac{\beta_i b_i}{\beta_i (1\!+\!
b_ip_i)\!+\!(1\!-\!\beta_i)(1\!+\!b_i p_i)^2}\!  \right) \!=\! 0, \;
\nonumber \\
& &  \qquad  \qquad \qquad \qquad \qquad \qquad \quad i=1, \cdots, N \\
& & \label{eq:NKKT4}   \frac{\beta_i b_i}{\beta_i
(1\!+\!b_ip_i)\!+\!(1\!-\!\beta_i)(1\!+\!b_i p_i)^2}\! \le \beta_i
{\nu }^\prime+{{\gamma }_i}^\prime,
\nonumber \\
& & \qquad  \qquad \qquad \quad \qquad  \qquad \qquad  i=1,
\cdots, N
\\
& & \label{eq:NKKT3} {0\le \ p}_i\ \le {p_i}^{max}, \quad i=1,
\cdots, N \\
& & \label{eq:NKKT6} \sum^N_{i=1}\beta_i{p_i} = p_R
\end{eqnarray}
Note that \eqref{NewNNeq:lagrange1} is rewritten as
(\ref{eq:NKKT4}) since $\lambda_i$ or, equivalently,
$\lambda_i^\prime$ is positive. Moreover, \eqref{eq:NKKT5} is
obtained by solving  \eqref{NewNNeq:lagrange1} with respect to
$\lambda_i^{\prime}$ and then substituting the result in
$\lambda_i^\prime p_i=0$. The following result gives a closed-form
solution for the system (\ref{eq:NKKT1})--(\ref{eq:NKKT6}), and
thus, it also gives an approximate solution for the power
allocation optimization problem \eqref{obj}--\eqref{eq:problem}.

\textit{\bf Theorem 2:} {\it For a set of DF relays, the
approximate power allocation $\{ p_1\ ,\dots ,\ p_N \}$, i.e., the
approximate solution of the optimization problem
(\ref{obj})-(\ref{eq:problem}), is}
\begin{equation} \label{eq:optsol}
p_i= \left( \frac{ -\beta_i + \sqrt{\beta_i^2 + 4 (1 - \beta_i)
b_i / {\nu}^\prime }}{2 b_i (1 - \beta_i)} - \frac{1}{b_i}
\right)_{0}^{p_i^{max}}
\end{equation}
{\it where ${\nu }^\prime$ is determined so that $\sum_{i=1}^N
\beta_i p_i = p_R$.}

See the proof in Appendix C.

It is interesting to mention that the power allocation scheme of
\cite{Arash} is a special case of the power allocation method
given by \eqref{eq:optsol} when ${\beta_i \rightarrow 1}$. Indeed,
it can be checked that for ${\beta_i \rightarrow 1}$, we have
\begin{eqnarray}
p_i \!\!\!&=&\!\!\! \left(\lim_{\beta_i \rightarrow 1} \frac{-\!
\beta_i \!+\! \sqrt{\beta_i^2 \!+\! 4 (1 \!-\! \beta_i) b_i /
{\nu}^\prime }}{2b_i (1 \!-\! \beta_i)} \!-\! \frac{1}{b_i}
\right)_{0}^{p_i^{max}} \nonumber \\
\!\!\!&=&\!\!\! \left(\frac{1}{\nu^\prime} - \frac{1}{b_i}
\right)_{0}^{p_i^{max}}
\end{eqnarray}
This special case shows that the total power should be distributed
only among relays from a selected set and the relays with `better'
channel conditions use bigger portion of the total power. Note
that this solution can be viewed as a form of water-filling
solution and has similar complexity. However, the solution
\eqref{eq:optsol} is not interpretable only by the means of
water-filling, and the optimal power allocation among the relays
and relay admission depend on the average probability of correct
decoding by relay nodes and also on the corresponding average
channel gain-to-noise ratios for admitted relays.

\section{Simulation results}
We consider a cooperative relay network consisting of a
source-destination pair and $N=5$ DF relay nodes.  In the first
phase, source transmits its message to the destination and the
relay nodes, while in the second phase, only the relay nodes,
which decoded the source message correctly, retransmit it to the
destination. All the nodes use QPSK modulation for data
transmission and noise power is assumed to be equal to $1$. Relays
are assumed to be located on a line of unit length connecting the
source and the destination nodes. Positions of the relay nodes with
respect to the source node are then randomly selected according to
the uniform distribution and are $ \{0.0117, 0.1365,0.2844,
0.4692, 0.8938 \}$. All the channel coefficients are modeled as
zero mean complex Gaussian random variables. The variance of the
channel coefficient between $p$th and $q$th nodes is
$1/d_{p,q}^\nu$ where $d_{p,q}$ is the distance between the nodes
and $ 2\leq \nu \leq 6$ is the path-loss parameter which is
assumed to be equal to $3$ throughout our simulations.

\subsection{Closed-Form SEP}
We first aim at comparing the closed form SEP expression
\eqref{eq2} with the SEP obtained through Monte-Carlo simulations
and demonstrating their equivalence. The source node transmits its
message to the destination node through only the relay nodes
$1,3,5$ or all relay nodes $1,2,3,4,5$. The total power is equally
divided among the source and the relay nodes. Fig.~1 shows the
average SEP corresponding to the closed-form expression
\eqref{eq2} and the SEP found by Monte-Carlo simulations with each
point obtained by averaging over $10^8$ independent runs. The
corresponding SEPs are shown versus the total power. Fig.~1
confirms the fact that the closed-form expression \eqref{eq2}
results in the same average SEP as the one obtained through
numerical simulations. The closed-form-based SEP and the SEP
obtained through numerical simulations coincide in both scenarios
considered with 3 and 5 relay nodes.
\begin{figure}[d]
\begin{center}
\includegraphics[scale=.45]{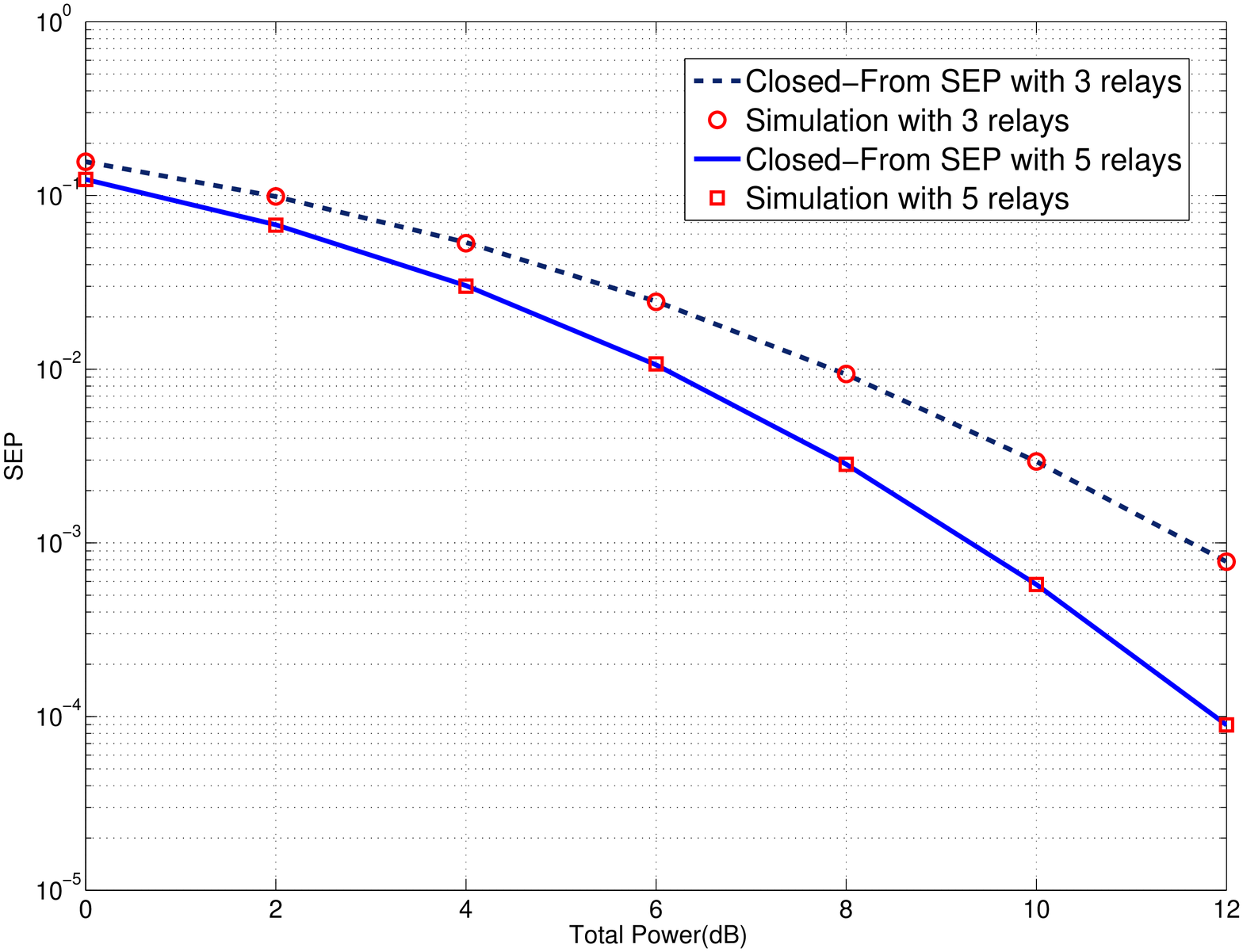}
\caption{Average SEP obtained through Monte-Carlo simulations and
by using the closed form SEP expression \eqref{eq2}.}
\end{center}
\end{figure}

\subsection{Accuracy of the Approximate Closed-Form Solution}
In this simulation example, we study the accuracy of the proposed
approximate power allocation method (see Theorem~2) by comparing
its performance to that of the optimal SEP found by solving the
exact problem \eqref{obj}--\eqref{eq:problem} using the algorithm
summarized in Appendix~B. Moreover, the performance of the
proposed power allocation strategy is compared with that of the
equal power allocation based strategy. The source transmit power
is assumed to be fixed and equals to $1$. The power transmitted by
each relay node in the scheme with equal power allocation is $p_i=
p_R/ \sum_{i=1}^N{\beta_i}, i=1,\ldots,R$. For evaluating the SEP,
the closed form the SEP expression in \eqref{eq2} is used.

\begin{figure}[t]
\begin{center}
\includegraphics[scale=.45]{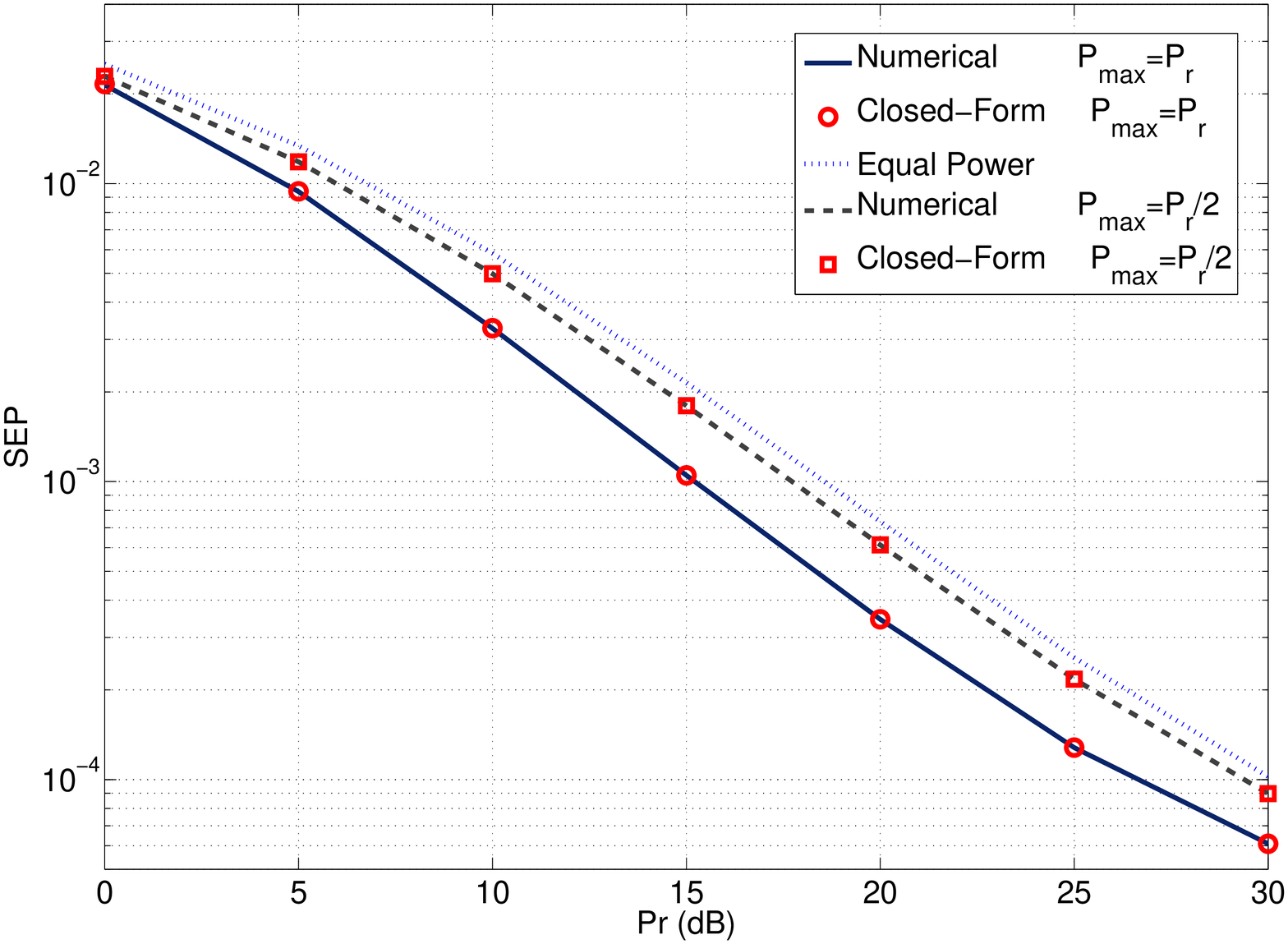}
\caption{Average SEP obtained based on the proposed exact and
approximate power allocation methods and the equal power
allocation method with 3 relays.} \label{Fig1}
\end{center}
\end{figure}

\begin{figure}[t]
\begin{center}
\includegraphics[scale=.45]{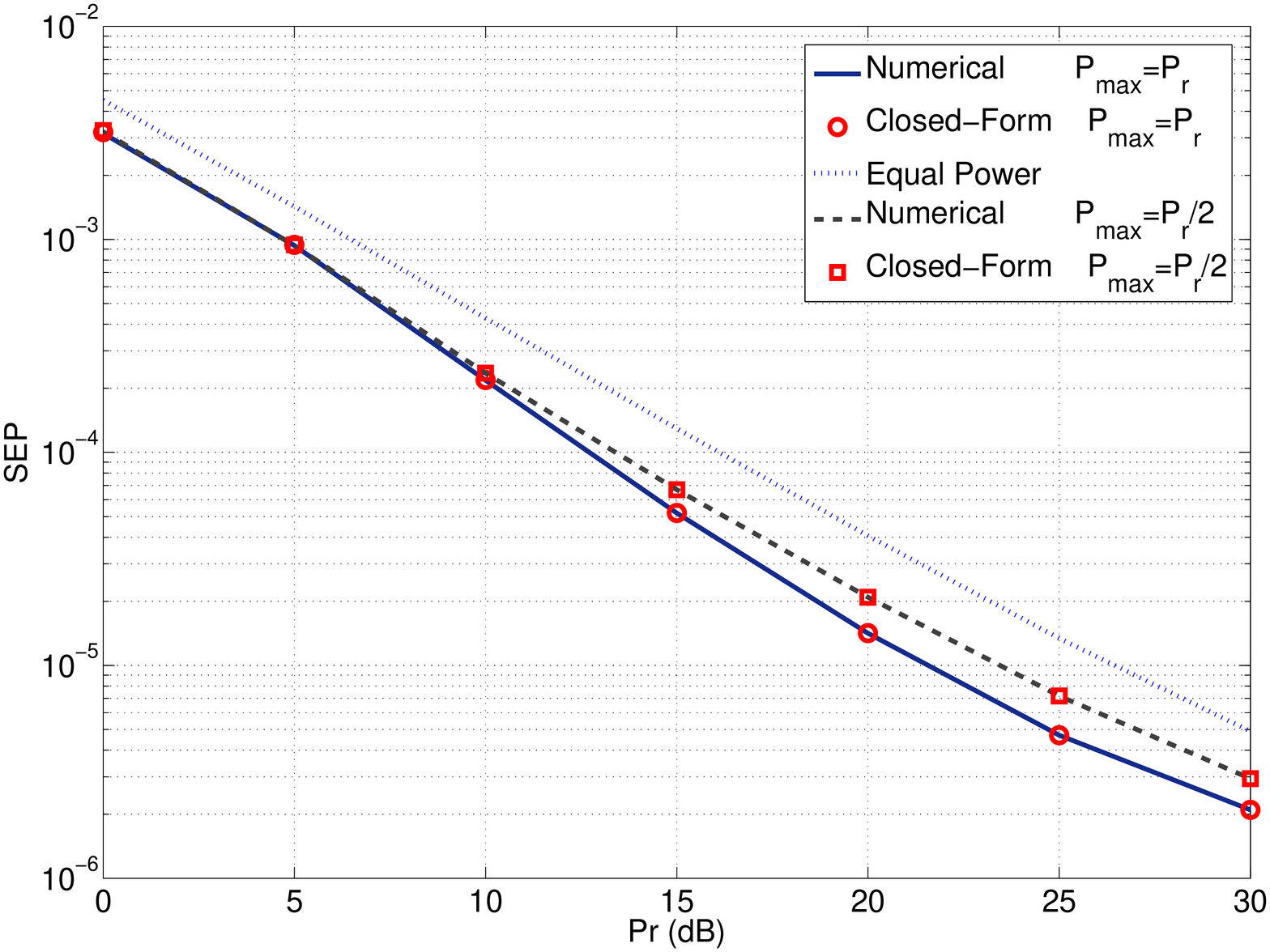}
\caption{Average SEP obtained based on the proposed exact and
approximate power allocation methods and the equal power
allocation method with 5 relays.} \label{Fig2}
\end{center}
\end{figure}

Figs.~\ref{Fig1}~and~\ref{Fig2} illustrate the SEP of the
aforementioned power allocation methods when only the relays
$1,3,5$ or all relays $1,2,3,4,5$ are used for the cases when
$p_{i}^{max}=p_r, i=1, \ldots,N$ and $p_{i}^{max}=p_r/2, i=1,
\ldots,N$. From these figures, it can be observed that the SEP
corresponding to the power allocation obtained by approximating
KKT conditions \eqref{eq:optsol} is very close to the optimal SEP
obtained through numerical solution of the optimization problem
\eqref{obj}--\eqref{eq:problem} even in the case when there are
only 3 relay nodes. It is noteworthy to mention that our extensive
simulation results confirm the high accuracy of the proposed
approximate power allocation method for different number of relay
nodes. 
It can be observed from Figs.~\ref{Fig1}~and~\ref{Fig2} that there
is more than $5$~dB performance improvement for the proposed
optimal power allocation scheme as compared to the equal power
allocation scheme in high SNRs and when, $p_i^{max}=p_r$. As it is
expected, if maximum available/allowable power at the relay nodes
is limited, the corresponding performance improvement
deteriorates.

\subsection{Performance Improvement Compared to Recent Methods}
In our last simulation example, we aim at comparing the
performance of our proposed power allocation scheme with another
recently proposed power allocation scheme. The other scheme aims
at minimizing the power required by all AF relay nodes under the
constraint that the average SEP does not exceed a certain desired
value and it uses the knowledge of the average channel gains
\cite{Maham}. Thus, as compared to our proposed power allocation
scheme, which minimizes the average SEP of a DF cooperative
network subject to a fixed total power of the relays nodes, the
method of \cite{Maham} minimizes the total power required by the
relay nodes of an AF cooperative network and fix the bound on the
average SEP. Both schemes use only the knowledge of the average
channel gains.

Compared to the method of \cite{Maham} according to which all
relays always retransmit the source message, our proposed method
exploits the additional information about the source message
decoding failure/success in each relay node, and thus, not all
relays retransmit the message. As a result, it is expected that
the proposed method will outperform the one in \cite{Maham}. In
addition, an approximation of SEP, which is accurate only at high
SNRs, is used in \cite{Maham}, while the exact SEP expression is
derived and used for our power allocation scheme. To ensure a fair
comparison between two aforementioned power allocation schemes, we
first find optimal power allocation according to the scheme of
\cite{Maham}. Then, we use the total power obtained by the scheme
of \cite{Maham} as a power bound for our proposed power allocation
scheme to ensure that the network operates with the same amount of
total power in both cases.

\begin{figure}[t]
\begin{center}
\includegraphics[scale=.45]{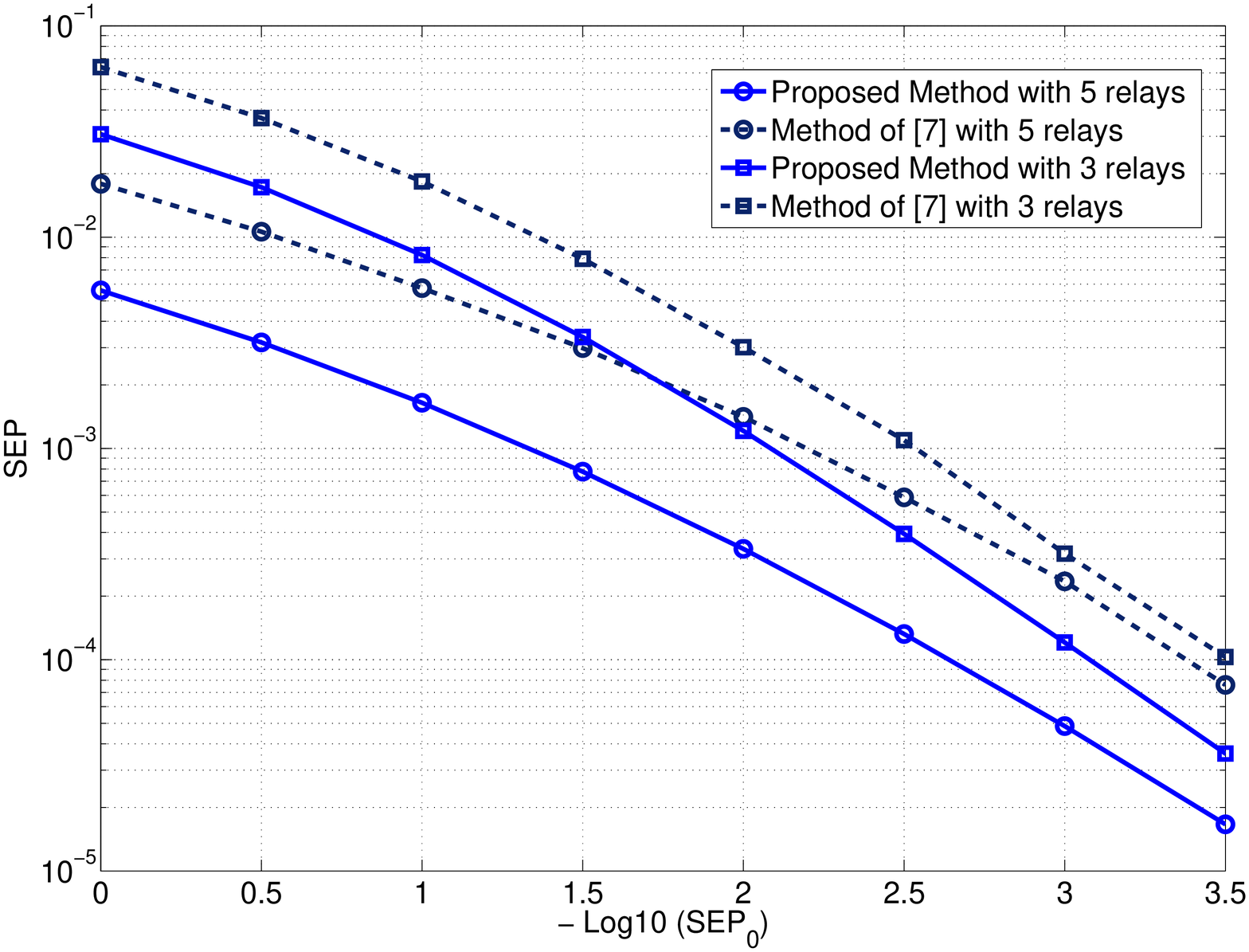}
\caption{Average SEP obtained based on the proposed approximate
power allocation method and the method of \cite{Maham} using $3$
and $5$ relays.} \label{PC3}
\end{center}
\end{figure}

Fig.~\ref{PC3} compares the SEP corresponding to the optimal power
allocation of the AF cooperative network using \cite{Maham} and
the SEP corresponding to the optimal power allocation obtained
based on the proposed method for DF networks for relays $1, 3, 5$
and all relays $1, 2, 3, 4, 5$, respectively, versus different
desired SEPs denoted as SEP$_0$ for the case that there is no
restriction on the maximum allowed power for each relay node.
Monte-Carlo simulations with $10^6$ independent runs are used for
obtaining each point in Fig.~\ref{PC3}. It can be observed from
this figure that, as expected, the proposed power allocation
scheme has superior performance over that of the scheme of
\cite{Maham} with the same average transmit power from the relay
node. This improvement can be attributed to the fact that in the
proposed scheme the relay for which the channel condition between
the source and relay is not good enough or, equivalently, the
relay that can not decode the source message correctly, does not
transmit. This additional information is exploited in the proposed
scheme, while it is not used in the scheme of \cite{Maham}. In
addition, the exact average SEP is used for the proposed scheme
versus the approximate one in the scheme of \cite{Maham}.

\section{Conclusion}
A new simple closed form expression for the average SEP for DF
cooperative relay network has been derived. Using this expression,
a new power allocation scheme for DF relay networks has been
developed by minimizing the average SEP under the constraints on
the total power of all relays and the maximum powers of individual
relays. The proposed scheme requires only the knowledge of the
average of the channel gains from relays to destination and the
direct link. The exact and approximate closed form solutions to
the corresponding optimization problem have been found and a high
accuracy of an approximate solution is demonstrated. According to
the proposed approximate solution, the power allocation and relay
admission depend on the average probability of correct decoding by
relay nodes and also on the corresponding average channel
gain-to-noise ratios for admitted relays. The improved performance
of the proposed scheme compared to some other schemes is
convincingly shown via simulations.

\section*{Appendix A}
Let us consider the following formula for the average SEP
\begin{eqnarray} \label{Pconvexity}
P_{e} \!\!\!&=&\!\!\!  \frac{1}{\pi} \sum_{k=0}^{2^N-1}
\bigg(\prod_{\{i|\boldsymbol \phi_k(i)=1\} \cup \{i=0\}} \!
\beta_i \cdot \prod_{\{i|\boldsymbol \phi_k(i)=0\}} \! (1 \!-\!
\beta_i) \nonumber \\
\!\!\!&\times&\!\!\! \int_0^{\frac{M-1}{M}\pi} \! \prod_{\{i |
\boldsymbol \phi_k (i) = 1\} \cup \{i=0\}}
\frac{\sin^2(\theta)}{\sin^2(\theta) \!+\! b_i p_i }d\theta \bigg).
\end{eqnarray}
The Hessian of the integral inside \eqref{Pconvexity}  with
respect to $\mathbf p$ can be obtained as
\begin{eqnarray}
\mathbf H \!\!\!&=&\!\!\! \frac{1}{\pi} \int_0^{\frac{M-1}{M}\pi}
\bigg(\mathbf a_{\boldsymbol \phi(k)} ( \theta, \mathbf p) \mathbf
a_{\boldsymbol \phi(k)}^H (\theta, \mathbf p) \nonumber \\
\!\!\!&+&\!\!\! \mathbf D_{\boldsymbol \phi(k)} ( \theta, \mathbf
p)) g_{\boldsymbol \phi(k)} (\theta, \mathbf{p}) \bigg) \, d\theta
\label{Hessian}
\end{eqnarray}
where
\begin{equation}
g_{{\boldsymbol \phi}_k} (\theta, \mathbf{p}) = \prod_{ \{i |
\boldsymbol \phi_k (i) = 1\} \cup \{i=0\}}
\frac{\sin^2(\theta)}{\sin^2(\theta) + b_i p_i }
\end{equation}
\begin{equation}
\mathbf a_{\boldsymbol \phi_k}(\theta, \mathbf p) \triangleq
\left( \frac{b_{1}\boldsymbol \phi_k (1)}{\sin^2 (\theta) + b_{1}
p_{1}}, \cdots, \frac{b_{N} \boldsymbol \phi_k (N)}{\sin^2
(\theta) + b_{N} p_{N}} \right)^T
\end{equation}
\begin{eqnarray}
\!\!\!& &\!\!\! \mathbf D_{\boldsymbol \phi(k)} (\theta,
\mathbf p) \!\triangleq \nonumber \\
\!\!\!& &\!\!\! diag \left( \! \frac{{b_{1}}^2\boldsymbol \phi_k
(1)}{(\sin^2 (\theta) \!\!+\!\! b_{1} p_{1})^2}, \!\cdots,\!
\frac{{b_N}^{2} \boldsymbol \phi_k (N)}{(\sin^2 (\theta) \!\!+\!\!
b_{N} p_{N})^2} \! \right)
\end{eqnarray}
with $( \cdot )^H$ denoting the Hermitian transpose and
$diag(\cdot)$ standing for a diagonal matrix.

Since the matrices $\mathbf a_{\boldsymbol \phi_k}(\theta, \mathbf
p) \mathbf a_{\boldsymbol \phi_k}^H (\theta, \mathbf p)$ and
$\mathbf D_{\boldsymbol \phi_k}(\theta, \mathbf p)$ are both
positive semi-definite for $\theta \in \left( 0, (M-1) \pi / M
\right)$ and $ p_i \geq 0$, $\ i=1,\cdots,N$ and also
$g_{\boldsymbol \phi_k} (\theta, \mathbf{p}) \geq 0$, the Hessian
matrix is positive semi-definite. The average SEP
\eqref{Pconvexity} is a linear combination of integral expressions
that are convex, and as a result, it is convex on nonnegative
orthant. Moreover, since the constraints of
(\ref{obj})--(\ref{eq:problem}) are linear and form a convex set
which is a subset of nonnegative orthant, the problem
(\ref{obj})--(\ref{eq:problem}) overall is convex. $\hfill\square$

\section*{Appendix B}
The numerical procedure for solving the problem
(\ref{obj})--(\ref{eq:problem}) is based on the interior-point
methods. Specifically, the \emph{barrier function method}, which
is one of the widely used interior-point methods, is applied. The
barrier function method for solving the problem
(\ref{obj})--(\ref{eq:problem}) can be summarized in terms of the
following algorithm.

\ \ \ Given strictly feasible ${\mathbf p}= (p_1,p_2,...,p_N)$,
$t=t^{(0)}$, $\mu >1$, and $\epsilon > 0$, where $t>0$ is the step
parameter of the barrier function method, $\mu$ is the step size
of the algorithm, and $\epsilon$ is the allowed duality gap
(accuracy parameter) (see \cite{IEEEhowto:Boyd}), do the following. \\
\ \ \ 1. Compute $ \mathbf p(t) \!=\! \arg \min \left\{
\frac{t}{\pi} \! \int_0^{\frac{M-1}{M}\pi} g(\theta, \mathbf p )
\, d\theta \!-\! \sum_{i=1}^{N}{\log{p_i}} \right.$  $ \left.
-\sum_{i=1}^N{\log(p_i^{max}-p_i)} \right\}$ subject to $
\sum_{i=1}^N{\beta_i p_i}=p_R $ starting at current $\mathbf p$
using Newton's method. \\
\ \ \ 2.  Update $\mathbf p := \mathbf p (t)$. \\
\ \ \ 3.  Stopping criterion: quit if  ${N}/{t} < \epsilon $\\
\ \ \ 4. Increase $t =\mu t$ and go to step~1.

In this algorithm, $\mathbf p(t)$ is called the central point and
the first step of the algorithm is called the centering step.
Then, at each iteration the central point {$\mathbf p(t) $} is
recomputed using Newton's method until $N/t < \epsilon$, that
guarantees that the solution is found with accuracy $\epsilon$,
i.e., {$P_e(\mathbf p) - P_e(\mathbf  p_{opt}) < \epsilon$}.

For solving the optimization problem in the centering step, which
is a convex problem with a linear equality constraint, the
extended Newton's method is applied. It can be summarized as
follows.

\ \ \ Use $\mathbf p$ as an initial point and select error
tolerance used in the Newton's method $\epsilon_r$. \\
\ \ \ 1. Compute the Newton step $\Delta \mathbf p$ and decrement
$\lambda(\mathbf p)$.\\
\ \ \ 2. Stopping criterion: quit if $\lambda(\mathbf p)^2/2 <
\epsilon_r $. \\
\ \ \ 3. Line Search: Choose step size $s$ by backtracking line
search.\\
\ \ \ 4. Update $\mathbf p := \mathbf p + s \Delta p$ and go to
step~1.

The Newton's step $\Delta \mathbf p$ and decrement
$\lambda(\mathbf p)$ are obtained from the following equations
\begin{equation}
 \begin{pmatrix}
  \nabla^2 H(\mathbf p)  & \mathbf a  \\
  \mathbf a ^T  & 0
 \end{pmatrix}  \begin{pmatrix}
  \Delta \mathbf p   \\
  w
 \end{pmatrix}=  \begin{pmatrix}
  -\nabla H(\mathbf p)    \\
  0
 \end{pmatrix} \nonumber
\end{equation}
where
\begin{eqnarray}
h(\mathbf p) \!\!\!&=&\!\!\! \frac{t}{\pi} \int_0^{\frac{M-1}{M}
\pi} g(\theta, \mathbf p ) \, d\theta \nonumber \\
\!\!\!&-&\!\!\! \sum_{i=1}^{N}{\log{p_i}} - \sum_{i=1}^N
\log(p_i^{max}\!-\!p_i) \nonumber
\end{eqnarray}
and $\nabla^2 H(\mathbf p)$ and $\nabla H(\mathbf p)$ are the
Hessian and gradient of $H (\mathbf p)$, respectively. Finally,
decrement is defined as
\begin{equation}
\lambda(\mathbf p) =  (\Delta \mathbf p ^T \nabla^2 H(\mathbf p)
\Delta \mathbf p)^{1/2} \nonumber
\end{equation}
and $\mathbf a = (\beta_1, \ldots, \beta_N)$.

\section*{Appendix C}
First, note that the Lagrange multiplier $\nu^\prime$  is
non-negative, otherwise the equation \eqref{eq:NKKT4} implies that
$\gamma_i^\prime,\ i=1, \cdots, N$  are all positive numbers and,
in turn, equation \eqref{eq:NKKT2} implies that $p_i=p_i^{max}$,
$i=1, \ldots, N$. Since it was assumed that $\sum^N_{i=1} \beta_i
{p_i^{max}} > p_R$, the condition \eqref{eq:NKKT6} can not be
satisfied for negative $\nu^\prime$. As a result, $\nu^\prime$
must be non-negative. Depending on whether $b_i$ is greater or
smaller than ${\nu }^\prime$, two cases are possible. If $b_i\le
{\nu }^\prime$, then the condition (\ref{eq:NKKT5}) holds true
only if $p_i=0$. Indeed, if $b_i \le {\nu }^\prime$, the
expression $\beta_i {\nu }^\prime + {{\gamma }_i}^\prime - \beta_i
b_i / (\beta_i (1 + b_i p_i) + (1 - \beta_i) (1 + b_i p_i)^2)$
equals to the non-negative quantity $\beta_i {\nu }^\prime +
{{\gamma}_i}^\prime - \beta_i b_i$ when $p_i = 0$. Furthermore, by
considering the fact that the term $\beta_i b_i / \left( \beta_i
(1 + b_ip_i) + (1 - \beta_i) (1 + b_i p_i)^2 \right)$ is strictly
decreasing with respect to $p_i$, it is resulted that the
expression $\beta_i {\nu }^\prime+{{\gamma }_i}^\prime - \beta_i
b_i / (\beta_i(1 + b_ip_i) + (1 - \beta_i) (1 + b_i p_i)^2) $ is
greater than zero if $p_i >0 $, which means that the condition
(\ref{eq:NKKT5}) can not be satisfied if $p_i > 0$. Thus, $p_i$ is
zero when ${b}_i\le {\nu }^\prime$. Furthermore, if $b_i > {\nu
}^\prime$, then $p_i$ can not be equal to 0. It is because if
$p_i$ equals zero then the condition (\ref{eq:NKKT2}) implies that
${{\gamma }_i}^\prime=0$. Then, substituting $p_i = 0$ and
${{\gamma }_i}^\prime=0$ in the condition (\ref{eq:NKKT4}) yields
that $b_i \le {\nu }^\prime$, which contradicts the condition ${\
b}_i > {\nu }^\prime$. Therefore, using the condition
(\ref{eq:NKKT5}) and the fact that $p_i$ must be positive if ${\
b}_i > {\nu }^\prime$ we can infer that
\begin{equation} \label{mdequation}
\beta_i {\nu }^\prime+{{\gamma }_i}^\prime -\frac{\beta_i b_i}
{\beta_i (1 + b_i p_i) + (1 - \beta_i) (1 + b_i p_i)^2}\ \!=\! 0
\end{equation}

By solving \eqref{mdequation} with respect to $p_i$, we can find
that
\begin{eqnarray} \label{EI2}
{\ p}_i \!\!\!&=&\!\!\! \frac{1}{b_i} \left( \frac{-\beta_i \!+\!
\sqrt{\beta_i^2 \!+\! 4\beta_i (1 \!-\! \beta_i)b_i/({\beta_i
\nu}^\prime \!+\! {\gamma }_i^\prime)}}{2(1 \!-\! \beta_i)} \!-\!1
\right) \nonumber \\
\!\!\!& \triangleq &\!\!\! f_i(\nu^\prime, {\gamma }_i^\prime) .
\end{eqnarray}
Note that the negative root is not considered because $p_i$ must
be non-negative. We also defined above the function $f_i(\nu
^\prime, {\gamma }_i^\prime)$ for notation simplicity. By
substituting the latter expression for ${\ p}_i$ into
(\ref{eq:NKKT2}) and (\ref{eq:NKKT3}), we obtain
\begin{eqnarray}
& & \label{eq:NNKKT1} {{\gamma }_i}^\prime\left(f_i(\nu
^\prime,{\gamma}_i^\prime)-{p_i}^{max}\right)=0 \\
& & \label{eq:NNKKT2} 0\le f_i(\nu ^\prime, {\gamma }_i^\prime)
\le {p_i}^{max}.
\end{eqnarray}
If $f_i(\nu ^\prime, 0) > {p_i}^{max}$, then the function $f_i(\nu
^\prime, {\gamma }_i^\prime)$ is strictly decreasing with respect
to ${\gamma }_i^\prime$ since $\nu ^\prime$ is non-negative. Thus,
the condition (\ref{eq:NNKKT2}) holds true only if ${\gamma
}_i^\prime > 0$. In addition, the condition (\ref{eq:NNKKT1})
implies that $p_i = {p_i}^{max}$. Then the only remaining case is
when $0 \leq f_i(\nu ^\prime, 0) \le {p_i}^{max}$. In this case,
the conditions (\ref{eq:NNKKT1}) and (\ref{eq:NNKKT2}) hold true
only if ${{\gamma }_i}^\prime = 0$. Considering the fact that the
case of ${b}_i \le {\nu }^\prime$ is equivalent to the case when
$f_i(\nu ^\prime, 0) \leq 0$, we can conclude that all possible
cases are analyzed and they then can be summarized as
\eqref{eq:optsol}. \hfill $\square$


\begin{thebibliography}{1}

\bibitem{IEEEhowto:Survey} Y.~W.~Hong, W.~J.~Huang, F.~H.~Chiu,
and C.~C.~J.~Kuo, ``Cooperative communications in
resource-constrained wireless networks,'' {\it IEEE Signal
Porcessing Mag.}, vol.~24, no.~3, pp.~47–-57, May~2007.

\bibitem{IEEEhowto:Lane} J.~N.~Laneman, D.~N.~C.~Tse, and
G.~W.~Wornell, ``Cooperative diversity in wireless networks:
Efficient protocols and outage behavior,'' {\it IEEE Trans. Inf.
Theory}, vol.~50, no.~12, pp.~3062–-3080, Dec.~2004.

\bibitem{Sergiy}
K.~T.~Phan, T.~Le-Ngoc, S.~A.~Vorobyov, and C.~Tellambura, ``Power
allocation in wireless multi-user relay networks,'' {\it IEEE
Trans. Wireless Commun.}, vol.~8, no.~5, pp.~2535--2545, May~2009.

\bibitem{IEEEhowto:andrea} A.~Goldsmith, {\it Wireless Communications.}
Cambridge University Press: Cambridge, 2005.

\bibitem{Sergiy2} K.~T.~Phan, S.~A.~Vorobyov, and C.~Tellambura,
``Precoder design for space-time coded MIMO systems with
correlated Rayleigh fading channels using convex optimization,''
{\it IEEE Trans. Signal Processing},  vol.~57, no.~2,
pp.~814--819, Feb.~2009.

\bibitem{Yener}
M.~Chen, S.~Serbetli, and A.~Yener, ``Distributed power allocation
strategies for parallel relay networks,'' {\it IEEE Trans.
Wireless Commun.}, vol.~7, no.~2, pp.~552-–561, Feb. 2008.

\bibitem{Adve}
Y.~Zhao, R.~Adve, and T.~J.~Lim, ``Improving amplify-and-forward
relay networks: Optimal power allocation versus selection,'' {\it
IEEE Trans. Wireless Commun.}, vol.~6, no.~8, pp.~3114–-3123,
Aug.~2007.

\bibitem{Blets} A.~Bletsas, A.~Khisti, D.~P.~Reed, and A.~Lippman,
``A simple cooperative method based on network path
selection,''{\it IEEE Journal on Selected Areas in
Communications}, vol.~24, no. 3,pp.~659-–672, Mar.~2006.

\bibitem{Maric}
A.~Host-Madsen and J.~Zhang, ``Capacity bounds and power
allocation for wireless relay channels,''{\it IEEE Trans. Inform.
Theory}, vol.~51, no.~6, pp.~2020–-2040, Jun. 2005.


\bibitem{IEEEhowto:Luo} J.~Luo, R.~S.~Blum, L.~Cimini, L.~Greenstein,
and A.~Haimovich, ``Power allocation in a transmit diversity
system with mean channel gain information,'' {\it IEEE Commun.
Lett.}, vol.~9, pp.~616--618, Jul.~2005.

\bibitem{Maham} B.~Maham, A.~Hjorungnes, and M.~Debbah, ``Power
allocations in minimum-energy SER constrained cooperative
networks,'' {\it special issue on cognitive radio, Annals of
telecommunications - Annales des télécommunications}, vol.~64,
no.~7, pp.~545--555, Aug. 2009.

\bibitem{Sadek1}
W.~ Su, A.~K.~Sadek, K.~J.~R.~Liu, ``SER performance analysis and
optimum power allocation for decode-and-forward cooperation
protocol in wireless networks,'' in {\it Proc. IEEE Wireless
Communications and Networking Conf.}, New Orleans, LA, USA,
Mar.~2005, pp.~984--989.

\bibitem{Sadek2}
A.~K.~Sadek, W.~Su, K.~J.~R.~Liu, ``Multinode cooperative
communications in wireless networks," {\it IEEE Trans. on
Signal Processing,} vol.~55, no.~1, pp.~341--355, Jan.~2007.

\bibitem{Sou}
Y.~ Lee, M.~H.~Tsai, S.~L.~ Sou, ``Performance of
decode-and-forward cooperative communications with multiple
dual-hop relays over nakagami-m fading channels,'' {\it IEEE
Trans. on Wireless Commun.}, vol.~8, no.~6, pp.~2853--2859,
June 2009.

\bibitem{Arash}
A.~Khabbazibasmenj and S.~A.~Vorobyov, ``Power allocation in
decode-and-forward cooperative networks via SEP minimization,'' in
{\it Proc. 3rd IEEE CAMSAP}, Aruba, Dutch Antilles, Dec.~2009,
pp.~328--331.

\bibitem{RepResearch} P.~Vandewalle, J.~Kovacevic, and M.~Vetterli,
``Reproducible research in signal processing,'' {\it IEEE Signal
Process. Mag.}, vol.~26, no.~3, pp.~37–47, May~2009.

\bibitem{IEEEhowto:Alouini} M.~K.~Simon and M.-S.~Alouini,
``A unified approach to the performance analysis of digital
communication over generalized fading channels,'' {\it Proc.
IEEE}, vol.~86, pp.~1860–-1877, Sept.~1998.

\bibitem{IEEEhowto:Boyd} S.~Boyd and L.~Vandenberghe,
{\it Convex Optimization.} Cambridge University Press: Cambridge,
2004.

\end{thebibliography}
\end{document}